\documentclass[%
 reprint,
 superscriptaddress,
 amsmath,amssymb,
 aps,
]{revtex4-1}

\usepackage{graphicx}
\usepackage{dcolumn}
\usepackage{bm}
\usepackage{hyperref}
\usepackage{algorithm}
\usepackage[noend]{algpseudocode}

\begin{document}

\preprint{APS/123-QED}

\title{
	Quantitative shadowgraphy and proton radiography for large intensity modulations
	}

\author{Muhammad Firmansyah Kasim}
\affiliation{
 John Adams Institute, Denys Wilkinson Building, Keble Road, Oxford OX1 3RH, United Kingdom
}
\author{Luke Ceurvorst}
\author{Naren Ratan}
\author{James Sadler}
\author{Nicholas Chen}
\affiliation{
 Clarendon Laboratory, Department of Physics, University of Oxford, Oxford OX1 3PU, United Kingdom
}
\author{Alexander S\"{a}vert}
\affiliation{
 Institut f\"{u}r Optik und Quantenelektronik, Abbe-Center of Photonics, Friedrich-Schiller-Universit\"{a}t, 07743 Jena, Germany
}
\affiliation{
Helmholtz-Institut Jena, Friedrich-Schiller-Universit\"{a}t, 07743 Jena, Germany
}
\author{Raoul Trines}
\author{Robert Bingham}
\affiliation{
 STFC Rutherford Appleton Laboratory, Chilton, Didcot OX11 0QX, United Kingdom
}
\author{Philip N. Burrows}
\affiliation{
 John Adams Institute, Denys Wilkinson Building, Keble Road, Oxford OX1 3RH, United Kingdom
}
\author{Malte C. Kaluza}
\affiliation{
 Institut f\"{u}r Optik und Quantenelektronik, Abbe-Center of Photonics, Friedrich-Schiller-Universit\"{a}t, 07743 Jena, Germany
}
\affiliation{
Helmholtz-Institut Jena, Friedrich-Schiller-Universit\"{a}t, 07743 Jena, Germany
}
\author{Peter Norreys}
\affiliation{
 Clarendon Laboratory, Department of Physics, University of Oxford, Oxford OX1 3PU, United Kingdom
}
\affiliation{
 STFC Rutherford Appleton Laboratory, Chilton, Didcot OX11 0QX, United Kingdom
}
\date{\today}

\begin{abstract}
Shadowgraphy is a technique widely used to diagnose objects or systems in various fields in physics and engineering. In shadowgraphy, an optical beam is deflected by the object and then the intensity modulation is captured on a screen placed some distance away. However, retrieving quantitative information from the shadowgrams themselves is a challenging task because of the non-linear nature of the process. Here, a novel method to retrieve quantitative information from shadowgrams, based on computational geometry, is presented for the first time. This process can also be applied to proton radiography for electric and magnetic field diagnosis in high-energy-density plasmas and has been benchmarked using a toroidal magnetic field as the object, among others. It is shown that the method can accurately retrieve quantitative parameters with error bars less than 10\%, even when caustics are present. The method is also shown to be robust enough to process real experimental results with simple pre- and post-processing techniques. This adds a powerful new tool for research in various fields in engineering and physics for both techniques.
\end{abstract}

\pacs{Valid PACS appear here}
\maketitle

\section{Introduction}
Shadowgraphy is a technique to visualise modulations in discrete objects \cite{shadowgraphy-book, shadowgraphy-book2} and is used extensively in our daily life. For example, when sun rays propagate through a transparent object with non-flat surface, one can readily observe the modulation of the light intensity behind the object on a screen placed a suitable distance away. This occurs because when light rays propagate, different refractive indices in the object cause the rays' paths to be deflected, resulting in the intensity modulation. With its simplicity, the shadowgraphy technique has become a widely used diagnostic tool in many different fields in physics and engineering. Examples are diagnosing plasma wakefields \cite{savert-wakefield-shadowgraphy}, measuring temperatures in combustion processes \cite{lewis-temperature-measurement}, and characterization of optical systems \cite{paper-dari-jimmy}.
A similar technique, that of  proton radiography, is also widely employed to diagnose the structure in laser-plasma experiments \cite{nilson-magnetic-reconnection, borghesi-soliton, li-magnetic-reconnection, romagnani-channeling, willingale-channeling, huntington-weibel-instability}. In proton radiography, instead of using light rays, a proton beam is fired into the plasma. The electric and magnetic fields inside the plasma deflect the protons' trajectories. Proton beams are both highly laminar and have discrete divergence angles that allow magnification of the object, provided that the screen is placed far enough away from the object. By looking at the intensity modulation of the proton beam on the screen, one can see the structure inside the plasma with $\sim \mathrm{\mu m}$ resolution. Among the applications of proton radiography are studies of experimental magnetic reconnection phenomena \cite{nilson-magnetic-reconnection, li-magnetic-reconnection}, observing solitons \cite{borghesi-soliton}, laser channeling in plasmas \cite{romagnani-channeling, willingale-channeling}, and observing the Weibel instability \cite{huntington-weibel-instability}.

One emphasises here that both the shadowgraphy and proton radiography techniques share the same underlying principle. Thus, one can refer to proton radiography as shadowgraphy and vice-versa, without losing generalities.  We will do this throughout this paper. By doing so, we show that this new approach provides a powerful new quantitative diagnostic tool for high-energy-density plasma science. 

Although shadowgraphy is widely used in plasma science, in many cases it is used as a qualitative analysis tool \cite{romagnani-channeling, willingale-channeling}. There have been many efforts in the past to retrieve the quantitative information from shadowgrams, but it has only been possible, so far, in limited cases where the intensity modulations are small. This is done mainly by employing Poisson's equation solver \cite{shadowgraphy-book, pogany, quant-shadowgraphy-made-easy, kugland-invited-paper} or by using the diffusion equation \cite{paper-yang-baru-dan-bikin-kaget} for specific cases \cite{paper-dari-jimmy, bone-laser-plasma}. The equation for small intensity modulation of shadowgraphy was also obtained by Pogany, \textit{et al.} \cite{pogany} using phase contrast approach and Fresnel diffraction. The non-linear nature of shadowgraphy makes it a challenging task for large modulation cases. Some experiments also make use of a grid to estimate the deflection of the beam \cite{willingale-fast-advection, li-magnetic-reconnection}. However, the technique depends on the grid resolution and it becomes harder to estimate when the feature to be observed is about the same size as the grid resolution or smaller \cite{romagnani-channeling, willingale-channeling}.

In this paper a method to retrieve quantitative information from shadowgraphic images for large intensity modulations, without using a grid, is presented. A coherent beam for optical shadowgraphy is also assumed throughout. By retrieving the quantitative information one can interpret phenomena in much greater detail, and thus provide a greater understanding of the diagnosed system. Section \ref{sec:theory} provides equations underlying the shadowgraphy and proton radiography techniques as well as the basic tools used in the methods section. Then the new method to retrieve the quantitative information is explained, as well as its implementation, in section \ref{sec:method}. Benchmarking with simulations is presented in section \ref{sec:benchmark} and tests on real experimental results in section \ref{sec:experiment}. Section \ref{sec:conclusions} concludes the paper.

\section{Theory}\label{sec:theory}
\subsection{Deflectometry}
If beams of light or charged particles are fired into deflecting objects along the $z_0$-axis, they will be deflected by an amount of
\begin{equation}\label{eq:angle-deflection}
\mathbf{a} (x_0,y_0) = -\nabla \Phi(x_0, y_0),
\end{equation}
where $\Phi(x_0, y_0)$ is the deflection potential. The deflection potentials for optical shadowgraphy and proton radiography cases, respectively, are \cite{quant-shadowgraphy-made-easy, kugland-invited-paper} \begin{subequations}\label{eq:deflection-potentials}
\begin{align}
\Phi(x_0, y_0) &= -\int \ln \eta(x_0,y_0,z_0)\ \mathrm{d}z_0
\\
\Phi(x_0, y_0) &= \frac{q}{2W} \int \phi(x_0,y_0,z_0)\ \mathrm{d}z_0,
\\
\Phi(x_0, y_0) &= -\frac{q}{m} \int \mathbf{A} \cdot \ \mathrm{d}\mathbf{z_0}
\end{align}
\end{subequations}
where $\eta$ is the refractive index of the object in light shadowgraphy cases, $\phi$ and $\mathbf{A}$ respectively are the electric and magnetic potential, $q$, $W$, and $m$ are the charge, energy, and mass of the particle in the beam, respectively. It is assumed that the beam propagates in straight lines during the interaction with the object. 

With each deflection, beams at position $(x_0, y_0)$ on the object plane are mapped to position $(x, y)$ on the screen according the equations below,
\begin{equation}\label{eq:mapping}
\begin{array}{lr}
x = x_0 + \mathbf{a} \cdot \mathbf{\hat{x}}\ L  \\
y = y_0 + \mathbf{a} \cdot \mathbf{\hat{y}}\ L,
\end{array}
\end{equation}
where $L$ is the distance between the object and the screen.
These equations assume the beams are collimated before the interaction with the objects. For diverging beams using the paraxial approximation, one can simply replace $x_0 \rightarrow x_0 (1+L/l)$ and $y_0 \rightarrow y_0 (1+L/l)$, where $l$ is the distance from the beam source to the object.

From the mapping equations, one can obtain the intensity of the beam on the screen as \cite{kugland-invited-paper} \begin{equation}\label{eq:intensity-modulation}
I(x,y) = \frac{I_0 (x,y)}{\left|\frac{\partial(x,y)}{\partial(x_0,y_0)}\right|},
\end{equation}
where $I_0 (x,y)$ is the beam intensity on the screen without deflections.
The term $\left|\partial (x,y)/\partial (x_0,y_0)\right|$ is the determinant of the Jacobian matrix of $(x,y)$ with respect to $(x_0, y_0)$. The Jacobian in the denominator is what makes shadowgraphy cases non-linear for relatively large $\mathbf{a}$ or $L$. Moreover, if $\mathbf{a}$ or $L$ is large enough, it can make the determinant of the Jacobian matrix very small, hence it causes very high intensity at some positions on the screen. This is called \textit{caustic}.

It is assumed that the object does not emit or absorb the beam, so the total flux on the screen without the object (source profile) is the same as the total flux on the screen with the object (target profile). With this assumption, the problem can be restated as the Monge transport problem  \cite{monge}: how are the particles transported from the source profile to the target profile such that the total distance for all particles is minimised? This can be solved using a combination of Lloyd's algorithm \cite{lloyd}, Voronoi and power diagram \cite{aurenhammer-voronoi-diagram}, and optimization \cite{aurenhammer-minkowski-type}.

\subsection{Voronoi and power diagram}
\begin{figure*}
\includegraphics[scale=0.25]{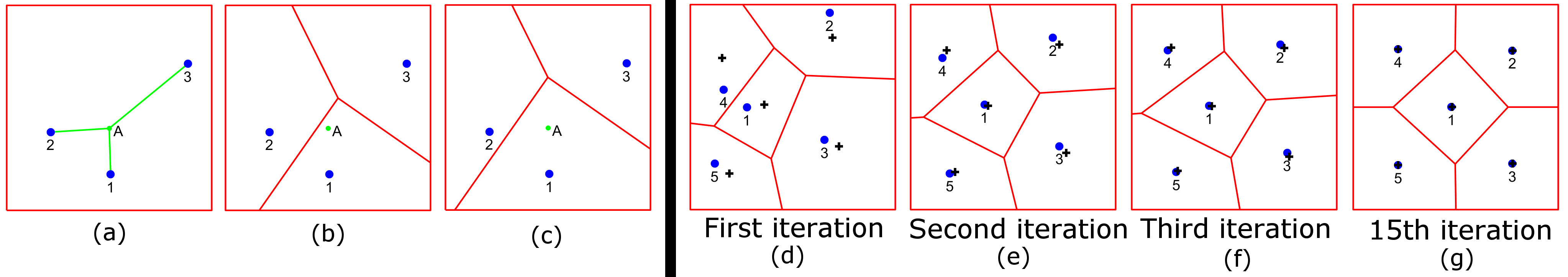}
\caption{\label{fig:complete-introduction}(a) An example of a case with 3 sites and a point A which is closest to the site 1. (b) A Voronoi diagram where the plane is divided into several regions based on which site is the closest one. (c) A power diagram with the same position of the sites as (b), but with more weight assigned to site 1. (d)-(g) Illustration of Lloyd's algorithm where the centroids of the regions are denoted by a plus (+) sign.}
\end{figure*}

Consider a 2D plane with several sites located on the plane. For every point on the plane, there is a site which is closest to the corresponding point. As an example, Figure \ref{fig:complete-introduction}(a) shows a plane with 3 sites and point A. Compared to the other sites, site 1 is the closest to the point A. Therefore point A belongs to site 1.

In the construction of a Voronoi diagram \cite{aurenhammer-voronoi-diagram}, the plane is divided by some regions. All points in a region belong to the site in the same region. Figure \ref{fig:complete-introduction}(b) is an example of a Voronoi diagram. Mathematically, the $i$-th site at $\mathbf{r}_{0i} = (x_{0i},y_{0i})$ occupies a region or cell on the source plane, $\mathbf{r}_0 = (x_0, y_0)$, where for all $j$,
\begin{equation}
|\!| \mathbf{r}_0 - \mathbf{r}_{0i} |\!|^2 \leq |\!| \mathbf{r}_0 - \mathbf{r}_{0j} |\!|^2.
\end{equation}

The equation above applies only for a case where all sites have the same weights. However, in some cases, this does not apply. A site with a larger weight tends to have a larger region compared to sites with smaller weights. A diagram resulting from weighted sites is called as weighted Voronoi diagram or power diagram. A region in power diagram is called as a \textit{power cell}. In the power diagram with weights $\mathbf{w}$, the $i$-th site at $\mathbf{r}_i = (x_i, y_i)$ occupies a region or power cell on plane $\mathbf{r} = (x, y)$ where
\begin{equation}
|\!| \mathbf{r} - \mathbf{r}_i |\!|^2 - w_i \leq |\!| \mathbf{r} - \mathbf{r}_j |\!|^2 - w_j
\end{equation}
for all $j$. Figure \ref{fig:complete-introduction}(c) shows an example of a power diagram with more weight on site 1. In a power diagram, it is possible for a site to not be located inside its region or even have no region. Setting all weights to be uniform or zero produces the Voronoi diagram.

\subsection{Lloyd's algorithm}
Lloyd's algorithm is a method of dividing a bounded plane into several regions with approximately the same area. The algorithm starts by deploying randomly a number of sites on a bounded plane. Then a Voronoi diagram is constructed to divide the plane into several regions. For every region, the algorithm calculates its centroid position. The sites are then moved to the centroid position of its region, and constructing the Voronoi diagram for the new positions. The process is then repeated until any stopping conditions are reached, e.g. maximum number of iterations, minimum displacement, etc. An illustration of the algorithm can be found on Figure \ref{fig:complete-introduction}(d)-(g).

There are some cases where the plane is not uniform. If this is the case, then there are several improvements that can be made. First, the site can be deployed randomly using a simple rejection method \cite{simple-rejection}. Positions on the plane with lower values tend to reject a site with higher probabilities. The rejected sites are deployed to other positions until they are accepted. Second, the centroid can be calculated by adjusting the values on the plane. It is similar to calculating the centre of mass of a 2D object with a non-uniform density.

\section{Method}\label{sec:method}

In order to retrieve quantitative information of the object from the screen, one needs the beam profiles both with and without the object in position. We refer to the beam profile without the object as the source profile, $I_0$, and the profile with the object as the target profile, $I$.

Initially, a number of sites are deployed randomly on the source plane profile with a simple rejection method mentioned above. Then, Lloyd's algorithm is applied on the source plane profile to distribute the sites so that each site has approximately the same flux. This produces a Voronoi diagram, or a power diagram with weights $\mathbf{w}=\mathbf{0}$. Once the Lloyd's algorithm finishes, then the algorithm performs optimization on the weights.

Denote $V_i$ as the $i$-th region on the source plane and $P_i^{\mathbf{w}}$ as the $i$-th region on the target plane as a function of all sites' assigned weights, $\mathbf{w}$. Note that $P_i^{\mathbf{0}} = V_i$. Also denote $S(V_i)$ and $T(P_i^{\mathbf{w}})$ as the flux of the $i$-th region on the source and target planes, respectively. The objective of the algorithm is to find the weights, such that transporting the flux from the source plane with intensity profile $I_0$ to the target plane produces the same intensity profile as the target profile, $I$, and the total distance travelled by all regions from the source plane to the target plane is minimised. Aurenhammer \cite{aurenhammer-minkowski-type} found that the weights can be found in the minimum of a convex function,
\begin{equation}\label{eq:convex-function}
f(\mathbf{w}) = -\sum_i \left[ w_i S(V_i) + \int_{P_i^{\mathbf{w}}} \left(|\!| \mathbf{r} - \mathbf{r}_{0i} |\!|^2 - w_i \right) I(\mathbf{r}) \mathrm{d}\mathbf{r} \right],
\end{equation}
where $\mathbf{r}_{0i}$ and $w_i$ are the $i$-th site position and the assigned weight, respectively.
It is noted that $\int_{P_i^{\mathbf{w}}} I(\mathbf{r}) \mathrm{d}\mathbf{r} = T(P_i^{\mathbf{w}})$. The gradient of the function is given by
\begin{equation}
\label{eq:gradient-of-weights}
\frac{\partial f(\mathbf{w})}{\partial w_i} = T(P^{\mathbf{w}}_i) - S(V_i),
\end{equation}
so any gradient based optimization methods can be employed. Note that in the optimization process, the sites positions do not change. It is only the assigned weights that are changed.
These equations have been employed to design surfaces of transparent objects that produce caustic designs \cite{schwartzburg-einstein-face}.

\begin{figure*}
\includegraphics[scale=0.15]{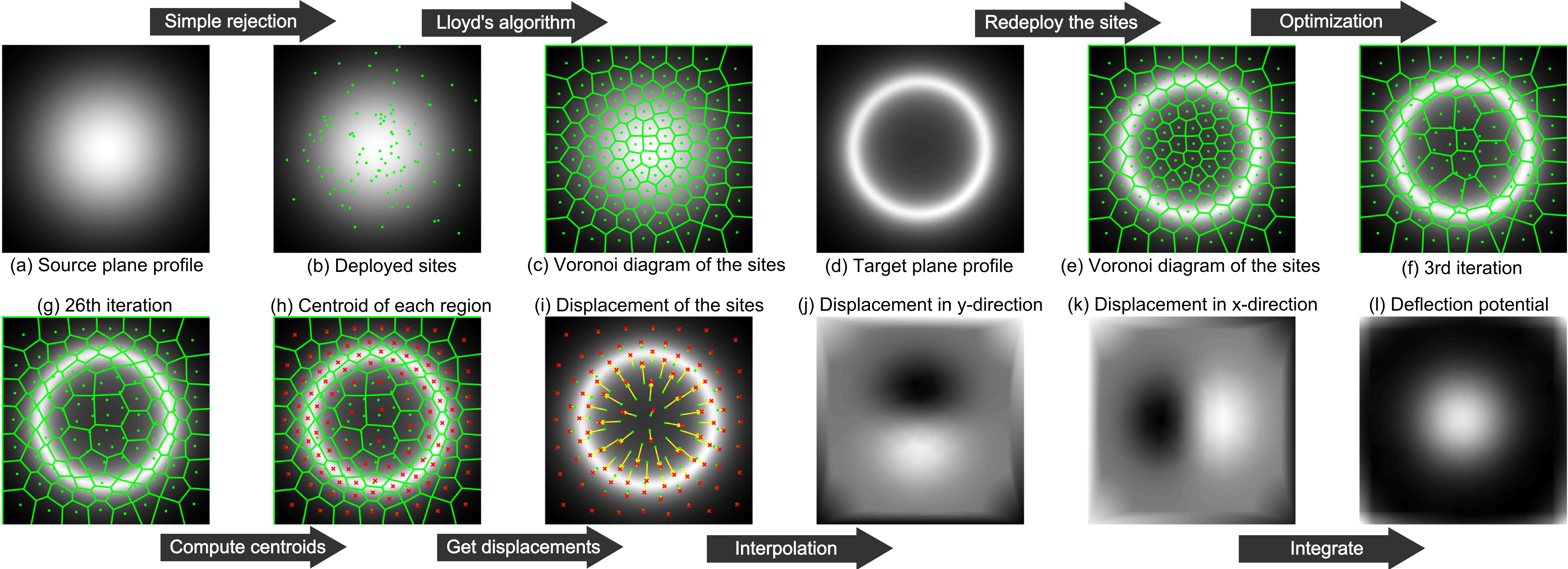}
\caption{\label{fig:algorithm-illustration} Illustration of the algorithm with 100 sites. Figure (b) shows the deployed sites in line \ref{alg-line:deploy} in Algorithm \ref{alg:inverse-shadowgraphy}. The results of Lloyd's algorithm from line \ref{alg-line:start-lloyds} to \ref{alg-line:finish-lloyds} is shown in Figure (c). The same sites positions are then redeployed on the target plane, as shown in Figure (e). Figures (f) and (g) are the results of line \ref{alg-line:optimization} of the algorithm on the $3^{\mathrm{rd}}$ and $26^{\mathrm{th}}$ iterations, respectively. Figure (h) and (i) are the results of lines \ref{alg-line:centroids} and \ref{alg-line:displacement}, respectively. The interpolated displacement in line \ref{alg-line:interpolation} is shown in Figure (j) and (k). Last, the integration of the displacement yields the deflection potential (l). The distortion at the corners in (l) is caused from moving 4 sites to the corner as in line \ref{alg-line:corners} in the algorithm.}
\end{figure*}

Once the minimum of equation \ref{eq:convex-function} is reached, the centroid position of each power cell in the power diagram, $ \mathbf{r_i} $, is computed. From the $i$-th power cell's centroid position on the target plane, $\mathbf{r_i}$, and the site's position on the source plane, $\mathbf{r_{0i}}$, the displacement in the $x$ and $y$ directions can be obtained by $\mathbf{a} = (\mathbf{r_i - r_{0i}})/L$. However, the displacement from source plane to target plane is obtained only at positions where the sites on the source plane are located. To fill in the displacement as a function of every position on the source plane, $\mathbf{a}(x_0,y_0)$, sites closest to the 4 corners are first moved to the corners and then the natural neighbour interpolation is used. The sites need to be moved to the corners so that the convex hull of the sites covers all the source plane and thus natural neighbour interpolation can be used. This causes some distortion near the corners, but this can be minimised by having more sites. The result of this method is curl-free at most positions, thus the deflection potential, $\Phi(x_0, y_0)$, can then be obtained by integrating the deflection in the $x$ or $y$ direction. We call this method `the power diagram method' in the remaining sections of this paper. The complete pseudocode of this algorithm is given in Algorithm \ref{alg:inverse-shadowgraphy} where all the bold face variables show the vectors of variables for all sites. The illustration is shown in Figure \ref{fig:algorithm-illustration}

\begin{algorithm}[H]
\caption{Inverse shadowgraphy and proton radiography}
\label{alg:inverse-shadowgraphy}
\begin{algorithmic}[1]
\State \textbf{Input:} a shadowgram or a proton radiogram image
\State \textbf{Output:} $\Phi$, the 2D deflection potential of the object
\State 
\State \% \textit{Initialisation}
\State Deploy sites randomly on the source plane, $\mathbf{x}$ and $\mathbf{y}$ \label{alg-line:deploy}
\Repeat \label{alg-line:start-lloyds}
	\State Construct the Voronoi diagram with sites at $\mathbf{x}$ and $\mathbf{y}$ on the source plane
	\State Calculate the centroid of each region, $\mathbf{x_c}$ and $\mathbf{y_c}$
	\State $\mathbf{x} \leftarrow \mathbf{x_c}$; $\mathbf{y} \leftarrow \mathbf{y_c}$
\Until{any stopping conditions reached}\label{alg-line:finish-lloyds}
\State Construct the Voronoi diagram with sites at $\mathbf{x}$ and $\mathbf{y}$ on the source plane \label{alg-line:voronoi-source}
\State Calculate $\mathbf{S}(\mathbf{V})$
\State
\State \% \textit{Gradient-based optimization}
\State $\mathbf{w} \leftarrow \mathbf{0}$
\Repeat 
	\State Construct the power diagram with $\mathbf{x}$, $\mathbf{y}$, and $\mathbf{w}$ on the target plane \label{alg-line:optimization}
	\State Calculate $\mathbf{T}(\mathbf{P^w})$ for each site
	\State Calculate $f(\mathbf{w})$ and $\Delta\mathbf{w} = \nabla_{\mathbf{w}}f(\mathbf{w})$
	\State Update $\mathbf{w} \leftarrow \mathbf{w} - \alpha \Delta\mathbf{w}$
\Until{any stopping conditions reached}
\State
\State \% \textit{Finalisation}
\State Construct the power diagram with $\mathbf{x}$, $\mathbf{y}$, and $\mathbf{w}$ on the target plane
\State Obtain the centroid positions, $\mathbf{x_P}$ and $\mathbf{y_P}$ \label{alg-line:centroids}
\State Assign the displacement, $\mathbf{x_P}-\mathbf{x}$ and $\mathbf{y_P}-\mathbf{y}$, to each site \label{alg-line:displacement}
\State Move 4 sites closest to the corners to the corners \label{alg-line:corners}
\State Get the displacement of each pixel using natural neighbour interpolation \label{alg-line:interpolation}
\State Integrate the displacement in $x$ or $y$ axis to obtain $\Phi$ \label{alg-line:result}

\end{algorithmic}
\end{algorithm}

\subsection{Implementation}
There are a lot of basic computational geometry algorithms employed in the implementation of this method. First, to obtain the power diagram of sites, algorithms that use convex hull and transformation to dual space are employed \cite{nocaj-power-diagram}. Voronoi diagram can be obtained by the same algorithm by setting all weights to zero. Bounded Voronoi and power diagrams inside a rectangle are obtained by clipping the diagram with the rectangle using the Sutherland-Hodgman algorithm \cite{sutherland-hodgman}. The Sutherland-Hodgman algorithm is employed for all polygon clippings in the implementation, since all polygons are convex in this case.

To calculate the function in equation \ref{eq:convex-function}, one needs to compute the weighted area (i.e. $S(V_i)$ and $T(P_i^{\mathbf{w}})$), weighted centroid position (i.e. $\mathbf{r_{0i}}$ and $\mathbf{r_{i}}$), and the weighted moment of inertia (i.e. $\int_{P_i^\mathbf{w}} |\!| \mathbf{r - r_{0i}} |\!|^2 I(\mathbf{r}) \mathrm{d}\mathbf{r}$) of each cell in the power diagram. In order to simplify the problem, it is reasonable to pixelate the intensity profile and assume the intensity within one pixel is constant. Thus, the above parameters can be computed by splitting the cell into several polygons with uniform density within a pixel, computing the parameters for each polygon, and merging the parameters to give the parameters for the given cell \cite{merigot}. The area, centroid position, and moment of inertia with respect to the origin of a 2D convex polygon with $N$ vertices can be shown to be
\begin{subequations}
\begin{align}
A = \frac{1}{2}\sum_{i=0}^{N-1} &(x_{i+1} y_i - x_i y_{i+1})
\\
x_c = \frac{1}{6A}\sum_{i=0}^{N-1} &(x_i + x_{i+1})(x_{i+1} y_i - x_i y_{i+1})
\\
y_c = \frac{1}{6A}\sum_{i=0}^{N-1} &(y_i + y_{i+1})(x_{i+1} y_i - x_i y_{i+1})
\\
I_z = \frac{1}{12} \sum_{i=0}^{N-1} &[(x_i^2+x_ix_{i+1}+x_{i+1}^2) + \\ 
&(y_i^2+y_i y_{i+1}+y_{i+1}^2)](x_{i+1} y_i - x_i y_{i+1}) \nonumber
\end{align}
\end{subequations}
where $(x_i, y_i)$ is the vertex position of each polygon and they are ordered in the clockwise direction. Note that $(x_{N}, y_{N}) = (x_0, y_0)$. The cells' centroids for Lloyd's algorithm are also computed by this method.

To obtain faster convergence to the global minimum of the function in equation \ref{eq:convex-function}, one can use a quasi-Newton gradient descent algorithm \cite{bfgs}. However, using a quasi-Newton algorithm requires $O(N_s^2)$ memory, where $N_s$ is the number of sites, and it can be very large computationally. Thus, using the limited memory BFGS (L-BFGS) method \cite{l-bfgs, l-bfgs-code} can save memory while still achieving fast convergence. One can also use a multi-stage approach to minimise equation \ref{eq:convex-function} faster \cite{merigot}. The complete implementation code of the algorithm on this paper can be found at \href{https://github.com/mfkasim91/invert-shadowgraphy}{https://github.com/mfkasim91/invert-shadowgraphy}.


\section{Benchmark with simulations}\label{sec:benchmark}
\subsection{Magnetic field proton radiography}
The first test for this method considers the case of a proton beam with energy of $W = 14.7\ \mathrm{MeV}$ propagating in the positive $z$-direction and going through a toroidal magnetic field. The toroidal magnetic field around the centre gives a line-integrated magnetic field on the object plane of
\begin{equation}
-\int\mathbf{B}\times \mathrm{d}\mathbf{z} = D_m \exp\left(-\frac{|\!|\mathbf{r}|\!|^2}{2\sigma^2}+\frac{1}{2}\right) \frac{1}{\sigma} \mathbf{r}
\end{equation}
where $D_m$ is the maximum value of line-integrated value of the toroidal magnetic field. This basic structure has been found in laser-plasma experiments, such as in magnetic reconnection experiments \cite{nilson-magnetic-reconnection, li-magnetic-reconnection}. Even though only magnetic field cases are considered here, it can be expanded into light shadowgraphy and electric field cases using equations \ref{eq:deflection-potentials}.

\begin{figure}
\includegraphics[scale=0.27]{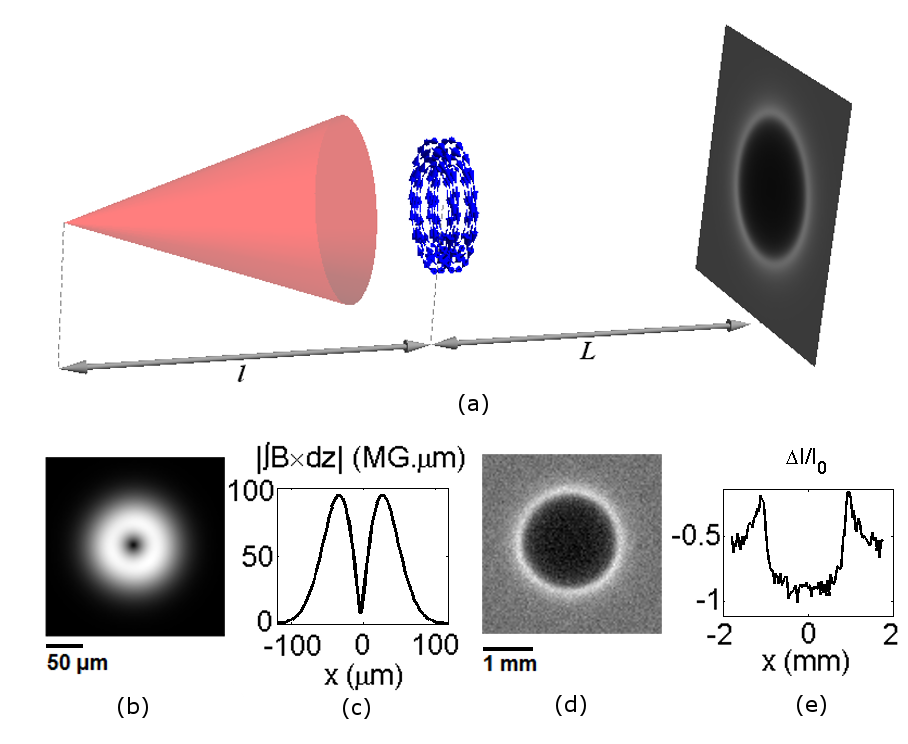}
\caption{\label{fig:toroidal-system} Illustration of the test case system. In figure (a), the proton beam  is fired through a toroidal magnetic field and the intensity modulation is captured on the screen. Figure (b) shows the magnitude of the line-integrated magnetic field, $-\int|\mathbf{B}\times\mathrm{d}\mathbf{z}|$ with maximum value, $D_m = 97\ \mathrm{MG\,\mu m}$, and its horizontal 1D-cross-section in figure (c). The intensity modulation on the screen is shown in figure (d) with its 1D-cross-section on figure (e). The intensity on the screen is augmented by Gaussian noise with variance 10\% of the average intensity value to test robustness of the method.}
\end{figure}

The transverse size of the toroidal magnetic field is assumed to be $\sigma = 30\ \mathrm{\mu m}$. The beam is deflected by the magnetic field and captured on the screen $L=2\ \mathrm{cm}$ away. The distance from the source to the magnetic field is $l=1.3\ \mathrm{mm}$, thus giving magnification of 15. It is assumed that the magnetic field extent in the $z$-direction is very small compared to $l$ and $L$. Visualisation of the test case can be seen in Fig. \ref{fig:toroidal-system}.

The beam's deflected velocity is $\mathbf{v_t} = -e/m \int \mathbf{B}\times \mathrm{d}\mathbf{z}$ where $e/m$ is the charge-to-mass ratio of the proton beam. Thus, the deflected angle is $\mathbf{a} = -e/\sqrt{2 m W} \int \mathbf{B}\times\mathrm{d}\mathbf{z}$. This gives the deflection potential as in equation \ref{eq:deflection-potentials}, given $\mathbf{B} = \nabla\times\mathbf{A}$.

\begin{figure}
\includegraphics[scale=0.27]{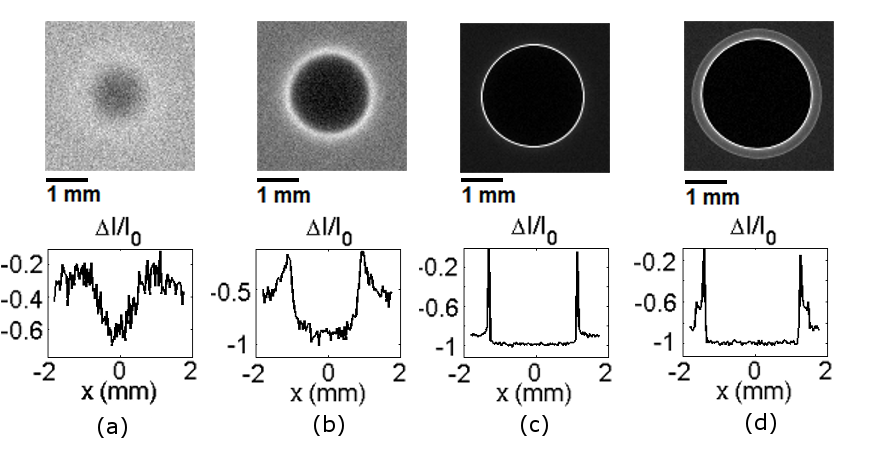}
\caption{\label{fig:toroidal-intensity-and-lineout}The beam's intensity modulation on the screen and its horizontal 1D-cross-section at the middle position for (a) small intensity modulation, (b) large intensity modulation, (c) a case where caustics have formed, and (d) where the caustics already branched. The value of $D_m$ for cases (a)-(d) is $(24, 97, 194, 290)\ \mathrm{MG\,\mu m}$, respectively.}
\end{figure}

The value of $D_m$ is varied from $10\ \mathrm{MG\,\mu m}$ to $340\ \mathrm{MG\,\mu m}$. These values cover the cases from small intensity modulation to the cases where caustics are formed. Caustics start to appear on the screen at $D_m = 190\ \mathrm{MG\,\mu m}$. The beam's intensity modulation is shown in Fig \ref{fig:toroidal-intensity-and-lineout}. Gaussian noise is added to each image with variance about $10\%$ of the average intensity.

From each image of the intensity modulation, the deflection potentials are retrieved using the method explained in this paper. Then one calculates the magnitude of the line integrated magnetic field, $|\!|\int \mathbf{B}\times\mathrm{d}\mathbf{z}|\!|$, from the deflection potential. The retrieved value is then compared with the original value to benchmark the method.

\begin{figure}
\includegraphics[scale=0.33]{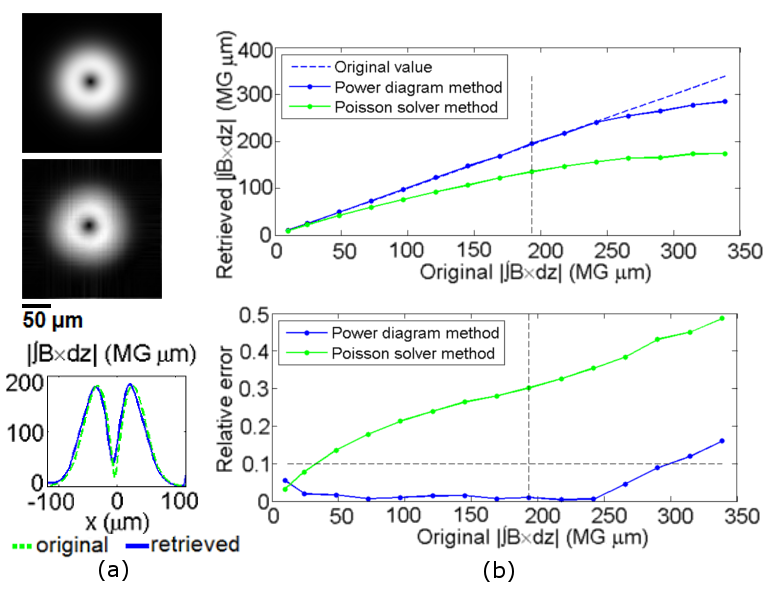}
\caption{\label{fig:toroidal-results} Comparison of the magnitude of the line-integrated magnetic field between (a,top) the original profile and (a,middle) the retrieved profile. The 1D-cross-section at the centre position of the original and the retrieved profile is given in (a,bottom). The dashed green line shows the 1D-cross-section of the original profile while the solid blue line shows the retrieved profile. The picture in (a) is taken for the case with $D_m = 194\ \mathrm{MG\,\mu m}$, where caustics are just formed. The quantitative comparison for the maximum value of the line-integrated magnetic field is given in figure (b). The top picture (b,top) shows the maximum retrieved value of $|\!|\int\mathbf{B}\times\mathrm{d}\mathbf{z}|\!|$ using the power diagram method (blue/dark grey) and the Poisson's equation solver method (green/light grey) compared with the original values. The relative error between the retrieved and original values is given in figure (b,bottom). The dashed vertical lines in figures (b) show the value when caustics are present. The dashed horizontal line shows the relative error of 10\%.}
\end{figure}

The images of the retrieved line-integrated magnetic field are shown in Fig \ref{fig:toroidal-results}(a). Comparison between the peaks of the retrieved values of line-integrated magnetic field and the original values are presented, as well as their relative errors. To see the improved performance of the method described in this paper, the line-integrated magnetic field profiles are retrieved using Poisson's equation solver and compared. Note that no noise has been added to the intensity images for the Poisson's equation solver case. These comparison results are shown in Fig \ref{fig:toroidal-results}(b).

It can be seen that the retrieved line integrated magnetic field gives very good agreement with the original profile, even when caustics are formed. The error on the retrieved value increases just when the branches of the caustics are relatively distinguishable. This is because in regions between the caustics branches, the beams are coming from more than one different position on the object plane, while this method assumes that each region on the target plane is formed from one region on the object plane only. However, one can still infer magnitude of the deflection potential in this case within some error.

A slightly higher relative error at small values of $|\!|\int\mathbf{B}\times\mathrm{d}\mathbf{z}|\!|$ is caused by the noise. The intensity modulation at that point is comparable to the noise. As the intensity modulation gets larger, the effect of noise seems to be weaker.

\begin{figure*}
\includegraphics[scale=0.3]{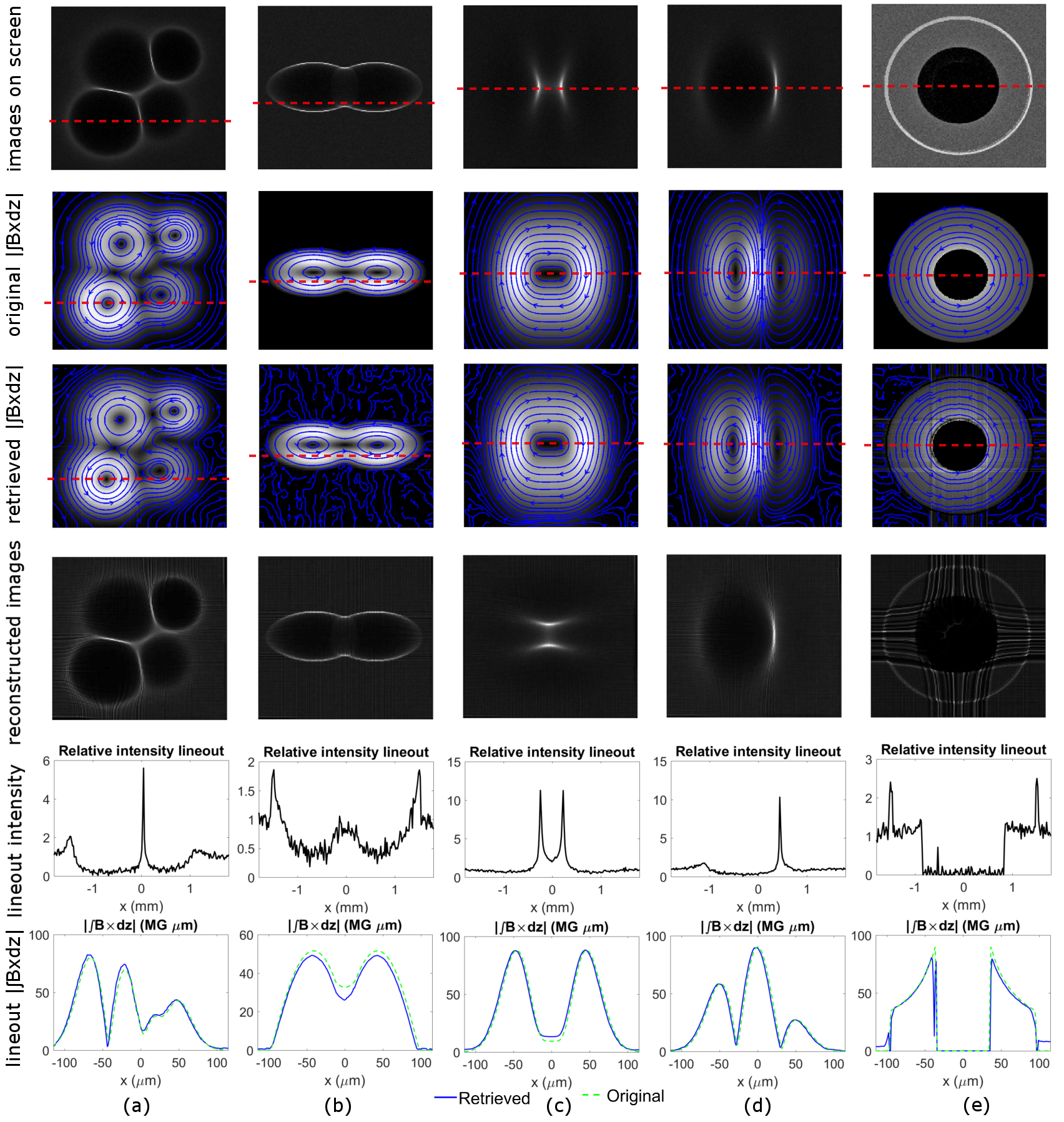}
\caption{\label{fig:benchmark-many-structures} Benchmark with arbitrary structures. (a) A case with 4 derivative Gaussian toroidal magnetic fields with different strengths and sizes in counter-clockwise direction. (b) Two toroidal magnetic fields where each potential has the form of $\cos^2$ with ellipsoidal shape. (c) Two close ellipsoidal Gaussian toroidal magnetic fields in clockwise direction, giving focusing effect on the proton beam. (d) Gaussian toroidal magnetic field with different directions, i.e. on the left it is counter-clockwise while on the right it is clockwise. (e) A case with two coaxial conductors with counter propagating currents between the two conductors. In this case, the conductors obstruct the beam. Magnitude of the magnetic field in each structure is set to form caustics on the target plane, except on (e) which shows a case with obstruction. The reconstructed images from the retrieved fields are presented to increase the confidence of the results. Gaussian noise with variance 10\% of the average intensity is also added to the images before the magnetic field information is retrieved. The 1D-cross-sections are taken at the positions indicated by dashed lines.}
\end{figure*}

It is observed that this method amplifies low-frequency components of the image and reduces the high-frequency components. It makes the power diagram method somewhat less robust to low frequency noise, but robust to high frequency noise.  This can be solved by applying a high frequency filter to the image before it is processed using the power diagram method.

On the other hand, the retrieved value using the Poisson's equation solver deviates significantly from the original value. The Poisson's equation solver gives relative error of 10\% when $D_m \gtrsim 30\ \mathrm{MG\,\mu m}$ while the power diagram method gives the same relative error when $D_m \gtrsim 300\ \mathrm{MG\,\mu m}$. If one wants an accuracy of less than 10\%, the power diagram method gives 10 times larger working range than the Poisson's equation solver's working range. 

As an additional benchmark to the case described above, we also tried retrieving quantitative information for arbitrary magnetic field structures. Fig \ref{fig:benchmark-many-structures} shows the retrieval results of the magnetic field with various structures with the same setup as the previous case. Each image has a size of $200\times200$ pixels with depth of 16 bits. The code was run on a highly parallel computer cluster using 32 cores. It takes around 2-3 hours to process one image.

In Fig \ref{fig:benchmark-many-structures}, one can see that the retrieved magnetic field structures agree very well with the original magnetic field. It is also apparent that the retrieved magnetic field can be different from what it seems in the proton radiography image. Moreover, the size of the structure can also be different, as shown in Fig \ref{fig:benchmark-many-structures}(e) where the structure's size is actually smaller than it seems in the proton radiography image.

\section{Tests with experimental results}\label{sec:experiment}
An analysis method will only be useful if it can be shown to work on real experimental data and give reasonable results. In this section, the power diagram method is used to analyse experimental data from S\"{a}vert, \textit{et al.} in cases of plasma wakefield shadowgraphy \cite{savert-wakefield-shadowgraphy}. In the experiment, a laser pulse was fired into a plasma to generate an electron density modulation wave associated with a laser-driven wakefield. Another laser pulse with much lower intensity was fired perpendicularly to the wakefield as a probe for the shadowgraphy method. The electron density fluctuations of the wakefield caused local modulations of the refractive index in the plasma. The refractive index modulation in the plasma caused the probe's path to bend so that some parts of the probe were brighter than others at the detector. The refractive index of a plasma with density profile $n(x_0,y_0,z_0)$ is 
\begin{equation}\label{eq:plasma-refractive-index}
\eta(x_0,y_0,z_0) = \sqrt{1-\frac{n(x_0,y_0,z_0) e^2}{m\epsilon_0\omega^2}},
\end{equation}
where $\epsilon_0$ is the vacuum permittivity constant, $e$ and $m$ are the electron's charge and mass, respectively, and $\omega$ is the frequency of the light. Using equations \ref{eq:deflection-potentials} and \ref{eq:plasma-refractive-index} with approximation $\omega_p^2 = n_0 e^2/m\epsilon_0 \ll \omega^2$, the deflection potential for light in a plasma is
\begin{equation}\label{eq:deflection-potential-plasma}
\Phi(x_0,y_0) \approx \frac{e^2}{m\epsilon_0\omega^2} \int n(x_0,y_0,z_0)\ \mathrm{d}z.
\end{equation}
Thus, the information that can be retrieved from shadowgrams using the power diagram method is $\int n\ \mathrm{d}z$.

One of the main challenges in inverting the shadowgrams for real experimental data is the non-uniformity of the probe's unmodulated intensity, i.e. the intensity profile without deflection from objects. This can be a big problem because the inversion processes from shadowgrams to deflection potentials amplify low frequency components. Even though this can be solved by taking the intensity profile without the object, the data is not usually available or reliable because of shot-to-shot variations. Therefore, a straightforward solution is to apply a high-pass filter to either the shadowgrams and/or the resulting deflection potentials.

\begin{figure*}
\includegraphics[scale=0.27]{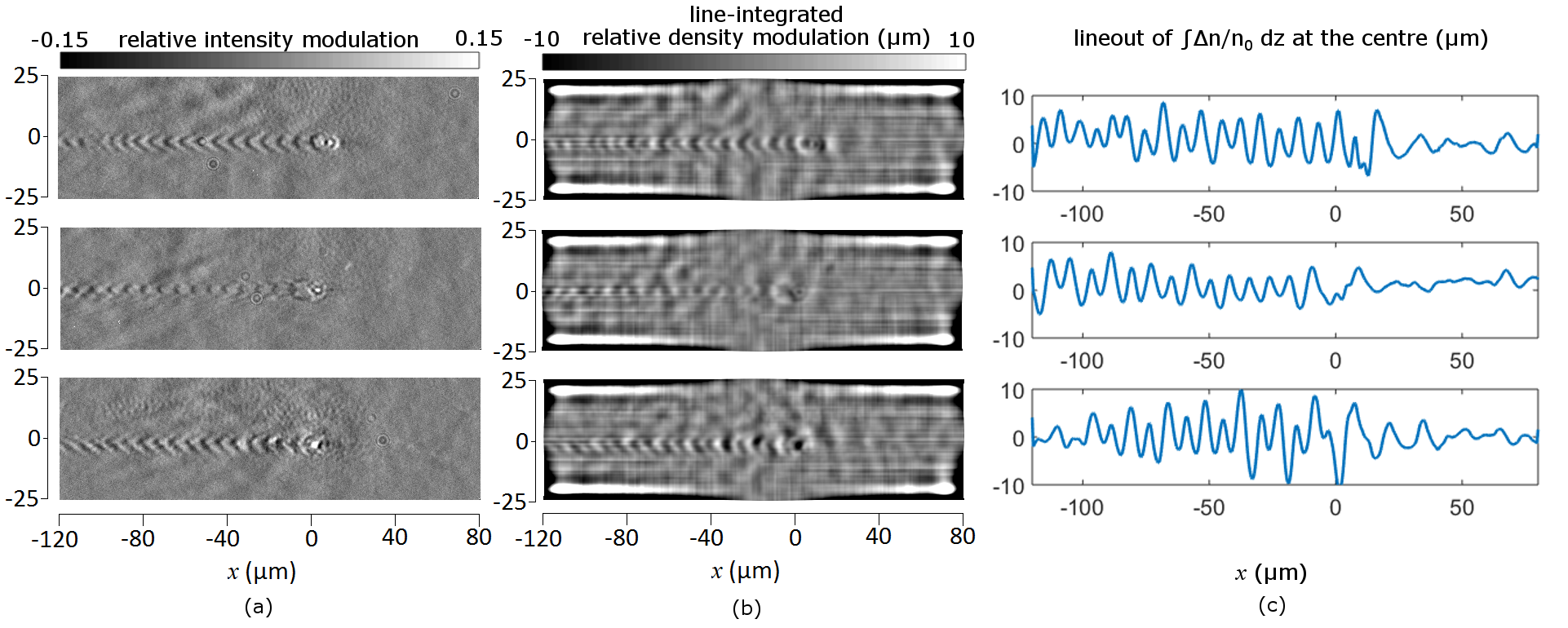}
\caption{\label{fig:plasma-shadowgraphy-experiment} (a) Original shadowgrams of plasma wakefield from S\"{a}vert, \textit{et al.}. Images reproduced with permission. (b) The retrieved line-integrated relative electron density modulation, $\int \Delta n/n_0\ \mathrm{d}z$ from the shadowgrams with their 1D-cross-sections at the centre of the wakefields in (c). High pass filter is applied in preprocessing and post-processing of the images. The very bright and very dark on the images are features obtained from applying high pass filter, as well as horizontal fringes on the left and right edges.}
\end{figure*}

The pulse that drives the wakefield has a wavelength of 810 nm, duration of 35 fs, and peak intensity $I_L = 6\times 10^{18}\ \mathrm{W\,cm}^{-2}$. The probe pulse has the same wavelength, but with shorter duration, 5.9 fs. The shadowgrams for a plasma with density $n_0 = 1.65\times10^{19}\ \mathrm{cm}^{-3}$ are shown in Fig. \ref{fig:plasma-shadowgraphy-experiment}(a). Using the power diagram method and equation \ref{eq:deflection-potential-plasma}, it is possible to infer the line-integrated relative electron density modulation, $\int \Delta n/n_0\ \mathrm{d}z$ from the shadowgrams. The inverted results from the shadowgrams are shown in Fig. \ref{fig:plasma-shadowgraphy-experiment}(b), where the grey scale shows the value of the line-integrated relative electron density modulation, $\int \Delta n/n_0\ \mathrm{d}z$, as well as their 1D-cross-sections at the centre of the wakefield in Fig. \ref{fig:plasma-shadowgraphy-experiment}(c). The figures still show the wakefield features with additional information of $\int \Delta n/n_0\ \mathrm{d}z$. It is shown from Fig. \ref{fig:plasma-shadowgraphy-experiment} that the power diagram method in this paper is robust enough to analyse real experimental results with additional preprocessing and post-processing. It should be noted that we have neglected relativistic effects in the plasma, e.g. the plasma electrons’ mass increase, which may occur during their interaction with the high-intensity driver pulses. To model this, multi-dimensional numerical simulations need to be applied \cite{e-siminos}

\section{Conclusions}\label{sec:conclusions}
We have presented a new method to retrieve quantitative data from shadowgraphic images. In the cases considered in this paper, a beam propagates through an object, gets deflected by it, and is then captured on a screen. The intensity modulation on the screen acts as the input and the deflection potential of the object is regarded as the output of this method. It assumes the beam propagates in straight lines while interacting with the object. Besides shadowgraphy, the method in this paper can also be applied to proton radiography cases.

The method has been benchmarked for a toroidal magnetic field case, which has been found in some laser plasma experiments, and a plasma wakefield shadowgraphy case. In some test cases, the method successfully retrieved the deflection potential profiles with relative error less than 10\% for large intensity modulation, even for cases where caustics are present. It is also tested using arbitrary structures of the diagnosed objects and gives very good results in retrieving structures with their quantitative parameters. Moreover, it has been shown that the method is also robust to noise, especially high-frequency noise. This extends the working range of the Poisson's solver equation by an order of magnitude. It is also shown that the method can be applied to real experimental results, with some additional pre-processing and post-processing. By applying this method, one can infer quantitative information from shadowgraphy images with high accuracy. This opens up a new dimension of research in a wide range of areas in engineering and physics.

\begin{acknowledgments}
The authors would like to acknowledge the support from the plasma physics HEC Consortium EPSRC grant number EP/L000237/1, as well as the Central Laser Facility and the Computer Science Department at the Rutherford Appleton Laboratory for the use of SCARF-LEXICON computer cluster. We would also like to thank ARCHER UK National Supercomputing Service for the use of the computing service. The authors would like to thank M. C. Levy for the useful discussions. We also wish to thank the UCLA/IST OSIRIS consortium for the use of OSIRIS. M.F.K. would like to gratefully thank Indonesian Endowment Fund for Education for its support. The authors gratefully acknowledge support from OxCHEDS and P.A.N. for his William Penney Fellowship with AWE plc. A.S. and M.C.K. are grateful for support by DFG (Grants No. TR18 B9, and No. KA 2869/2-1), and by BMBF (Contracts No. 05K10SJ2 and No. 03ZIK052).
\end{acknowledgments}


\begin{thebibliography}{99}
  \bibitem{shadowgraphy-book}
  G. S. Settles,
  \textit{Schlieren and Shadowgraph Techniques: Visualizing Phenomena in Transparent Media} (Heidelberg: Springer, 2001).
  
  \bibitem{shadowgraphy-book2}
   P. K. Panigrahi and K. Muralidhar, \textit{Schlieren and Shadowgraph Methods in Heat and Mass Transfer} (New York: Springer, 2012).
  
  \bibitem{savert-wakefield-shadowgraphy}
  A. S\"{a}vert, S. P. D. Mangles, M. Schnell, E. Siminos, J. M. Cole, M. Leier, M. Reuter, M. B. Schwab, M. Möller, K. Poder, O. J\"{a}ckel, \textit{et al.}, \href{http://dx.doi.org/10.1103/PhysRevLett.115.055002}{Phys. Rev. Lett. \textbf{115}, 055002 (2015)}.
  
  \bibitem{lewis-temperature-measurement}
  R. W. Lewis, R. E. Teets, J. A. Sell, and T. A. Seder, \href{http://dx.doi.org/10.1364/AO.26.003695}{Applied Optics Vol. 26, Issue 17, pp. 3695-3704 (1987)}.
  
  \bibitem{paper-dari-jimmy}
  H. Canabal, J. Alonso, and E. Bernabeu, \href{http://dx.doi.org/10.1117/1.1409939}{Opt. Eng. \textbf{40} (11) pp. 2517-2523 (2011)}.
  
  \bibitem{nilson-magnetic-reconnection}
    P. M. Nilson, L. Willingale, M. C. Kaluza, C. Kamperidis, S. Minardi, M. S. Wei, P. Fernandes, M. Notley, S. Bandyopadhyay, M. Sherlock, \textit{et al.}, \href{http://dx.doi.org/10.1103/PhysRevLett.97.255001}{Phys. Rev. Lett. \textbf{97}, 255001 (2006)}.
  
  \bibitem{li-magnetic-reconnection}
    C. K. Li, F. H. S\'{e}guin, J. A. Frenje, J. R. Rygg, R. D. Petrasso, R. P. J. Town, O. L. Landen, J. P. Knauer, and V. A. Smalyuk, \href{http://dx.doi.org/10.1103/PhysRevLett.99.055001}{Phys. Rev. Lett. \textbf{99}, 055001 (2007)}.

  \bibitem{borghesi-soliton}
  M. Borghesi, S. Bulanov, D. H. Campbell, R. J. Clarke, T. Zh. Esirkepov, M. Galimberti, L. A. Gizzi, A. J. MacKinnon, N. M. Naumova, F. Pegoraro, \textit{et al.}, \href{http://dx.doi.org/10.1103/PhysRevLett.88.135002}{Phys. Rev. Lett. \textbf{88}, 135002 (2002)}.
  
  \bibitem{romagnani-channeling}
  	L. Romagnani, A. Bigongiari, S. Kar, S. V. Bulanov, C. A. Cecchetti, T. Zh. Esirkepov, M. Galimberti, R. Jung, T. V. Liseykina, A. Macchi, \textit{et al.}, \href{http://dx.doi.org/10.1103/PhysRevLett.105.175002}{Phys. Rev. Lett. \textbf{105}, 175002 (2010)}.
  	
  \bibitem{willingale-channeling}
  	L. Willingale, P. M. Nilson, A. G. R. Thomas, J. Cobble, R. S. Craxton, A. Maksimchuk, P. A. Norreys, T. C. Sangster, R. H. H. Scott, C. Stoeckl, \textit{et al.}, \href{http://dx.doi.org/10.1103/PhysRevLett.106.105002}{Phys. Rev. Lett. \textbf{106}, 105002 (2011)}.
  
  \bibitem{huntington-weibel-instability}
  C. M. Huntington, F. Fiuza, J. S. Ross, A. B. Zylstra, R. P. Drake, D. H. Froula, G. Gregori, N. L. Kugland, C. C. Kuranz, M. C. Levy, \textit{et al.}, \href{http://dx.doi.org/10.1038/nphys3178}{Nat. Phys. \textbf{11}, 173–176 (2015)}.
  
  \bibitem{willingale-fast-advection}
  L. Willingale, A. G. R. Thomas, P. M. Nilson, M. C. Kaluza, S. Bandyopadhyay, A. E. Dangor, R. G. Evans, P. Fernandes, M. G. Haines, C. Kamperidis, \textit{et al.}, \href{http://dx.doi.org/10.1103/PhysRevLett.105.095001}{Phys. Rev. Lett. \textbf{105}, 095001 (2010)}.
  
  \bibitem{pogany}
  A. Pogany, D. Gao, and S. W. Wilkins, \href{http://dx.doi.org/10.1063/1.1148194}{Rev. Sci. Instrum. \textbf{68} (7) (1997)}.
  
  \bibitem{quant-shadowgraphy-made-easy}
  G. Izarra and C. Izarra, \href{http://dx.doi.org/10.1088/0143-0807/33/6/1821}{Eur. J. Phys. \textbf{33}, pp. 1821–1842 (2012)}.
  
  \bibitem{kugland-invited-paper}
  N. L. Kugland, D. D. Ryutov, C. Plechaty, J. S. Ross, and H.-S. Park, \textit{et al.}, \href{http://dx.doi.org/10.1063/1.4750234}{Rev. Sci. Instrum. \textbf{83}, 101301 (2012)}.

  \bibitem{paper-yang-baru-dan-bikin-kaget}
  C. Graziani, P. Tzeferacos, D. Q. Lamb, and C. Li, \href{http://arxiv.org/abs/1603.08617v1}{arXiv:1603.08617 [physics.plasm-ph] (2016)}.
  
  \bibitem{bone-laser-plasma}
   A. Bon\'{e}, N. Lemos, G. Figueira, and J. M. Dias, \href{http://dx.doi.org/10.1088/0022-3727/49/15/155204}{J. Phys. D: Appl. Phys. \textbf{49}, 155204 (2016)}.
   
 \bibitem{monge}
 G. Monge, \textit{Histoire de l'Acad\'{e}mie Royale des Sciences de Paris, avec les M\'{e}moires de Math\'{e}matique et de Physique pour la m\^{e}me ann\'{e}e}, pp. 666-704 (1781).
 
 \bibitem{lloyd}
 M. McCool and E. Fiume, \href{http://www.dgp.toronto.edu/~elf/.misc/poissondisk.pdf}{Proc. of the Graphics Interface, pp. 94-105 (1992)}. 
 
 \bibitem{aurenhammer-voronoi-diagram}
 F. Aurenhammer, \href{http://dx.doi.org/10.1145/116873.116880}{ACM Computing Surveys, Vol. 23, No. 3 (1991)}.
 
 \bibitem{aurenhammer-minkowski-type}
 F. Aurenhammer, F. Hoffmann, and B. Aronov, \href{http://dx.doi.org/10.1007/PL00009187}{Algorithmica, Vol. 20, Issue 1, pp. 61-76 (1998)}.

 \bibitem{simple-rejection}
 G. Casella, C. P. Robert, and M. T. Wells, \href{https://dx.doi.org/10.1214/lnms/1196285403}{Lecture Notes - Monograph Series, Vol. 45, pp. 342-347 (2004)}.

 \bibitem{schwartzburg-einstein-face}
 Y. Schwartzburg, R. Testuz, A. Tagliasacchi, and M. Pauly, \href{http://dx.doi.org/10.1145/2601097.2601200}{ACM Trans. Graph. 33, \textbf{4}, Article 74 (2014)}.
 
 \bibitem{nocaj-power-diagram}
 A. Nocaj and U. Brandes, \href{http://dx.doi.org/10.1111/j.1467-8659.2012.03078.x}{Computer Graphics Forum, Vol. 31, No. 3, pp. 855-864 (2012)}.
 
 \bibitem{sutherland-hodgman}
 I. E. Sutherland and G. W. Hodgman, \href{http://dx.doi.org/10.1145/360767.360802}{Communications of the ACM, Vol. 17, Issue 1, pp. 32-42 (1974)}.

 \bibitem{bfgs}
 A. G. Buckley, \href{http://dx.doi.org/10.1007/BF01609018}{Mathematical Programming, Vol. 15, Issue 1, pp. 200-210 (1978)}.
 
 \bibitem{l-bfgs}
 D. C. Liu and J. Nocedal, \href{http://dx.doi.org/10.1007/BF01589116}{Mathematical Programming, Vol. 45, Issue 1, pp. 503-528 (1989)}.
 
 \bibitem{l-bfgs-code}
 M. Schmidt. \href{http://www.cs.ubc.ca/~schmidtm/Software/minFunc.html}{minFunc: unconstrained differentiable multivariate optimization in Matlab (2005)}.

 \bibitem{merigot}
 Q. M\'{e}rigot, \href{http://dx.doi.org/10.1111/j.1467-8659.2011.02032.x}{Computer Graphics Forum, Vol. 30, Issue 5, pp. 1583-1592 (2011)}.
 
 \bibitem{e-siminos}
 E. Siminos, S. Skupin, A. S\"{a}vert, J. M. Cole, S. P. D. Mangles, and M. C. Kaluza, \href{http://dx.doi.org/10.1088/0741-3335/58/6/065004}{Plasma Physics and Controlled Fusion, Vol. 58, No. 6 (2015)}.
 
 
 
 
  	
 
  
  	
  
  
  	
  
  	
  
  
  
\end{thebibliography}
\end{document}